\documentstyle[aps,12pt,epsf]{revtex}
\def\la{\mathrel{\mathpalette\fun <}}
\def\ga{\mathrel{\mathpalette\fun >}}
\def\fun#1#2{\lower3.6pt\vbox{\baselineskip0pt\lineskip.9pt
  \ialign{$\mathsurround=0pt#1\hfil##\hfil$\crcr#2\crcr\sim\crcr}}}
\begin{document}

\draft
\title{Observing the Birth of Supermassive Black Holes with
the \lq\lq ICECUBE\rq\rq\ Neutrino Detector}

\author{Xiangdong~Shi, George~M.~Fuller}
\address{Department of Physics, University of California,
San Diego, La Jolla, California 92093-0319}

\author{and}
\author{Francis~Halzen}
\address{Department of Physics, University of Wisconsin,
Madison, Wisconsin 53705}

\date{March 16, 1998}

\maketitle

\begin{abstract}
It has been suggested that the supermassive black holes, at
the centers of galaxies and quasars, may initially form in
single collapses of relativistic star clusters or
supermassive stars built-up during the evolution of dense
star clusters. We show that it may be possible for ICECUBE
(a planned 1 km$^3$ neutrino detector in  Antarctica) to
detect the neutrino bursts associated with those collapses
at redshift $z\la 0.2$ with a rate of $\sim$ 0.1 to 1
burst per year. Such detections could give new insights
into the formation of structure in the universe,
especially when correlated with gravitational 
wave signatures or even gamma-ray bursts.
\end{abstract}
\bigskip


\newpage
In this letter we describe a possible means for detecting the
neutrino bursts accompanying the formation of supermassive
black holes at cosmological distances. (Hereafter the term \lq\lq
supermassive\rq\rq\ implies a mass $\ga 3\times 10^4\,M_\odot$.)
Mounting evidence suggests that supermassive black holes are
fairly common in the universe. Observations with the Hubble Space
Telescope point to supermassive black holes in nearly every
galaxy examined so far\cite{vanderMarel97}.
Likewise supermassive black holes have long been thought to be
the central engines of quasars and Active Galactic Nuclei
(AGN) \cite{rees84}. The masses inferred for these black holes
typically range from $\sim 10^6\,M_\odot$ (as in our galaxy)
to $\sim 10^9\,M_\odot$ (as in some quasars).
It is conceivable that every galaxy, large or small,
harbors a supermassive black hole at its center.  In other words,
supermassive black holes may have an abundance of one per
$10^{10}M_\odot$ (baryon mass) object,
the typical baryon content of dwarf galaxies such as M32.
Such a conjecture is entirely consistent with observations
if, like those nearby ones, most galactic nuclei are inactive
\cite{Kormendy}.

How these supermassive black holes were
formed is still a mystery.  However, a natural formation route
is through accretion onto a seed black hole. The seed black hole
presumably must be formed earlier from the collapse of a single
object or a cluster of objects. It is likely that these seed
black holes were themselves supermassive, having masses perhaps
only an order of magnitude or two below that of the current
end-state black holes. This is because it would be very
difficult for a much smaller seed black hole,
such as that resulting from an ordinary solar-mass-scale stellar
collapse, to accrete material efficiently enough so as
to reach a supermassive mass scale in a Hubble time at an
early epoch\cite{young77,loeb94}.

It has been suggested that the supermassive seed black holes may
form from collapses of dense star clusters or supermassive
stars that were built up as an intermediate phase during the final
evolutionary stage of collapsing star clusters \cite{begel78}.
Dense star clusters have been observed to reside at the center of
galaxies \cite{Eckart}, and in general have been
invoked to explain AGN activities \cite{Tal,Walter}.
The formation of an intermediate-stage supermassive star has
been argued to be on several of the possible evolution routes
of collapsing star clusters\cite{begel78}.

A supermassive star with mass $M\ga 3\times 10^4M_\odot$ is an
index $n=3$ polytropic configuration with a high entropy per
baryon $S\approx 300\,(M/10^5M_\odot)^{1/2}$
(in units of Boltzmann's constant)\cite{FWW}.
A supermassive star eventually collapses into a black hole as a
result of the Feynman-Chandrasekhar\cite{chandra64} instability.
Numerical calculations also have shown that a dense star
cluster of compact objects would eventually collapse as a 
result of the Feynman-Chandrasekhar instability just as would
a single supermassive object\cite{shapiro85}.

If a supermassive black hole formed as a result of a single
collapse, whether from a single supermassive star, or a
supermassive relativistic cluster of compact objects, 
we would expect the collapse to be accompanied by a burst
of thermally-produced neutrinos. The neutrino emission would
carry away a significant fraction of the gravitational binding
energy of the collapse, which is of order the rest mass energy
of the homologous core, $\sim 10^{59}M_5^{\rm HC}$ erg (where
homologous core mass, denoted by $(10^5M_\odot)M_5^{\rm HC}$,
is the mass that plunges through the event horizon as a unit).
The homologous core mass $M_5^{\rm HC}$ is only a fraction of the
initial stellar mass $M_5^{\rm init}\equiv M^{\rm init}/10^5M_\odot$,
with the fraction determined by the entropy loss during the
collapse \cite{shi98}.
This fraction is roughly $\sim$10\% for a non-rotational and
non-magnetized spherical supermassive star of $M_5^{\rm init}\sim 10$.

If the homologous core is massive enough
(namely $M_5^{\rm HC}\ga 0.1$), it will
be transparent to neutrinos\cite{shi98}.
This is in stark contrast to the case of
an ordinary core collapse supernova, where
the neutrinos are trapped in the core.
Since in our case the collapsing material
will be essentially in free fall, we
should have a set of well defined
neutrino luminosity/time templates with
which to compare to observations. Such 
time templates of neutrino luminosity
have a characteristic gradual rise
followed by a sudden drop due to the
core becoming a black hole.  Figure 1
shows an example of such a time profile
for the (un-normalized) neutrino 
luminosity, and the time profile of the
corresponding (un-normalized) signal in
ICECUBE. The duration of the neutrino 
burst is typically $\la t_{\rm dyn}$,
where $t_{\rm dyn}$ is the dynamic collapse 
timescale.  For a non-rotating, non-magnetized
progenitor, $t_{\rm dyn}\sim M^{\rm HC}_5$ seconds.

Shi and Fuller\cite{shi98} showed that for the collapse of
a non-rotating non-magnetized spherical supermassive object
the neutrino burst could carry away a fraction 
$\sim 0.04(M_5^{\rm HC})^{-1.5}$ of the total gravitational
binding energy (assuming $M_5^{\rm HC}\ga 0.1$, so that
this fraction is less than 1). 
The neutrino release is equally partitioned between neutrinos
and anti-neutrinos, with $\nu_e\bar\nu_e$ accounting for
70$\%$ of the total neutrino flux.
The average neutrino energy is similar for all species and
is $\approx 4(M_5^{\rm HC})^{-0.5}$ MeV.

If the progenitor is rapidly rotating or if magnetic stresses
are appreciable, the collapse timescale could increase
significantly. Neutrinos would then have a much better
chance of escaping before the core moves through an event
horizon. In such a case, the neutrino fluence could be an
order of magnitude higher than in the non-rotating case 
(but still limited by the total gravitational binding energy),
and the average neutrino energy could be a factor of 2 higher
as well\cite{shi98}.  The partition of energy among the
different neutrino species, however, would remain the same.

Though the neutrino emission accompanying
the collapse of a supermassive object is gigantic,
it is almost impossible to detect if it originates at a
redshift $z\ga 1$\cite{shi98}. But such a neutrino burst
may become detectable if the collapse occurs at a
redshift $z\la 0.2$. These recent supermassive collapse
events may not be uncommon, since there are still
large numbers of quasars (and AGNs) at redshift $\la 0.2$.

To estimate how many supermassive progenitors might collapse
at a redshift $z\la 0.2$, we need to know how abundant they 
are and how their collapse rate evolves with redshift.
A natural and reasonable starting point is to assume an
abundance of one per galaxy (including dwarf galaxies, i.e.,
one per $\sim 10^{10}M_\odot$ of baryons), and assume a collapse
rate that traces the formation rate of quasars, because 
quasars are powered by supermassive black holes. 
If we assume that the average quasar lifetime, which is
short ($\sim 10^8$ years) compared to the Hubble time,
does not evolve with redshift, the quasar formation rate
would then trace quasar number density as a function of redshift.
In figure 2 we replot the evolution of the quasar number
density (proper) as measured by Shaver {\sl et al.}
\cite{shaver98}, as well as that inferred from the quasar
catalogue compiled by V\'eron-Cetty and V\'eron\cite{veron}.
The plot assumes a Hubble constant 
$H_0=50$ km/sec/Mpc and a deceleration parameter $q_0=0.5$, 
which will be the working assumption in the following
calculations. (The details of cosmology do not affect our
results significantly.) The sample of Shaver {\sl et al.}
is radio-selected and is argued to have the least amount
of selection bias\cite{shaver98}.
We thus adopt the quasar number density evolution in the 
redshift range $z\ga 0.5$ as implied by this sample.
The catalogue of V\'eron-Cetty and V\'eron, on the other
hand, is a compilation of quasars brighter than absolute
magnitude $M_{\rm B}=-23$ from different samples with 
different selection criterion \cite{veron}.
Its degree of completeness is uncertain. Therefore, even
though the latter catalogue has far more data at the low
redshift end, its inferred quasar number density evolution
below $z\la 0.3$ cannot simply be taken as the extension of
the Shaver {\sl et al.}'s evolution function to $z\la 0.3$.
The catalogue does, however, seem to give a number density
evolution in fair agreement with Shaver {\sl et al.}'s
evolution function at $0.5\la z\la 2$. If nothing else,
the V\'eron-Cetty and V\'eron catalogue may indicate
that the quasar number density at $z\la 0.3$ potentially is
comparable to the quasar number density at $z\sim 0.5$.
To account for the uncertainty in quasar number density
evolution at the very low redshift, we will calculate the
nearby supermassive black hole formation rate for three
values of the relative quasar number density in the redshift
range $0.1\la z\la 0.2$ (the three thick solid lines in
figure 2), corresponding to a quasar number density that is
the same as (case 1), one-third of (case 2), and one-tenth
of (case 3) the quasar number density at $z\sim 0.5$.

We denote the collapse rate of supermassive objects as $\dot\phi$,
whose redshift evolution is inferred from figure 2.
Assuming a matter-dominated flat universe, we 
find that the total baryon mass in a unit volume that was
incorporated into supermassive black holes is now
\begin{equation}
\rho_{\rm sm}=\int_0^{t_0} {\dot\phi\,M\over (1+z)^3}\,{\rm d}t
\approx\,0.025\dot\phi_0\,M\,t_0,
\label{baryon1}
\end{equation}
where $t_0$ is the age of the universe today, $M$ is the mass of
the supermassive black hole today, and where $\dot\phi_0$ is the
normalization required to give the true formation rate in a unit
volume. On the other hand, we have assumed
$\rho_{\rm sm}=(M/10^{10}M_\odot)\rho_{\rm b}$ where
$\rho_{\rm b}\approx 1.88\times 10^{-29}\Omega_{\rm b}h^2$ g$\,$
cm$^{-3}$ (with $\Omega_b$ being the baryon density in
units of the critical density and $h\equiv H_0/(100 
{\rm km/sec/Mpc}$) is the baryon density today.  Therefore,
\begin{equation}
\dot\phi_0\approx
      7\times 10^{-17}\,\Bigl({\Omega_{\rm b}h^2\over 0.025}\Bigr)
       \Bigl({12.5{\rm Gyr}\over t_0}\Bigr)
       \,{\rm Mpc}^{-3}\,{\rm s}^{-1}.
\end{equation}

The event rate as observed now of these supermassive collapses
that occurred in the redshift range $z_1\le z\le z_2$ is
\begin{equation}
R(z_1,z_2)=\int_{z_1}^{z_2} \dot\phi\,
         {4\pi r^2\over (1+z)^4} {{\rm d}r\over {\rm d}z}{\rm d}z,
\label{rate}
\end{equation}
where $r$ is the comoving spatial coordinate.
Note that in this equation, in addition to the factor
$(1+z)^{-3}$ stemming from volume expansion, there is an
additional factor $(1+z)^{-1}$ due to the cosmic time dilation.
In a matter-dominated universe with $q_0=0.5$, we have
$r=3ct_0(1-1/\sqrt{1+z})$, where $c$ is the speed of light.
The rate of those collapses at $0.1\la z\la 0.2$ is then
\begin{equation}
R(0.1,0.2)=2\pi(3ct_0)^3\int_{0.1}^{0.2}\dot\phi\,
 {(1-1/\sqrt{1+z})^2\over (1+z)^{5.5}}{\rm d}z
 =R_0\,\Bigl({\Omega_{\rm b}h^2\over 0.025}\Bigr)
       \Bigl({12.5{\rm Gyr}\over t_0}\Bigr)\,\ {\rm year}^{-1},
\label{rate2}
\end{equation}
where $R_0\approx 1$, 0.3, and 0.1, for case 1, 2 and 3,
respectively. (The rate for all supermassive collapses is
roughtly 0.1 per day with the same scaling factors.)
These rates scale inversely with the amount
of baryon mass that contains one supermassive black hole,
which we have assumed to be $10^{10}M_\odot$ on average.
While we are certainly very interested in collapses at
$z\la 0.1$, we cut off the collapse rate at $z=0.1$ to
avoid extrapolating the quasar number density evolution
function too far toward low $z$.
Even so, the rate of nearby collapse of supermassive black
hole progenitors is potentially much higher than the Type II
(and Type Ib) supernova rate in our galaxy.

With these estimated characteristics, the neutrino bursts from
such relatively nearby supermassive object collapse events may
be detectable with ICECUBE, a $\sim 1$ km$^3$ neutrino detector
with $\sim 10^4$ optical modules soon to be proposed.
Just like the currently-operating AMANDA (Antarctic
Muon and Neutrino Detector Array) detector, it will
have sensitivity to bursts of neutrinos with energies
$\ga 10$ MeV\cite{halzen96}. The principle is similar
to that of detecting supernova neutrinos, i.e., to
detecting the {\v C}erenkov photon flashes resulting from
relativistic positrons produced by the reaction 
$\bar\nu_e\,+\,p\,\rightarrow\,n\,+\,e^+$.
The cross section for this process is 
$\approx 9\times 10^{-44}(E_{\bar\nu_e}/1\,{\rm MeV})^2$ cm$^2$.
The {\v C}erenkov photons are collected by optical modules
and their numbers are counted. Using the module efficiency 
and threshold estimates in Halzen, Jacobsen and Zas\cite{halzen96},
and the $\bar\nu_e$ spectrum calculated by Shi and Fuller\cite{shi98},
we find that the number of events expected in ICECUBE from a neutrino
burst resulting from the collapse of a supermassive object would be
\begin{equation}
N_{\rm event}\sim 10^{-3}\,N_M\,
\Bigl({750\,{\rm Mpc}\over d}\Bigr)^2\,
\Bigl({E_{\nu\bar\nu}\over
4\times 10^{57}(M_5^{\rm HC})^{-0.5}\,{\rm erg}}\Bigr)\,\alpha
\end{equation}
where $N_M$ is the number of optical modules in the detector,
$d$ is the proper distance to the source of the burst
($d\approx 750$ Mpc for $z=0.15$), 
$E_{\nu\bar\nu}$ is the total energy release in
neutrino emission, and $\alpha\sim{\cal O}(1)$
accounts for the energy spectrum of the $\bar\nu_e$ emission.
The expected number of $\bar\nu_e$ events per optical module
($N_{\rm event}/N_M$) from collapses of homologous cores with
various masses at $z=0.15$ is shown in Table 1.
In our calculation, we have imposed a 30$\%$ artificial 
limit on the fraction of the total gravitational binding
energy that can be carried away by neutrino emission. 
We believe this fraction is attainable, considering
that much higher efficiency is possible by magnetic
mechanisms.

Optimal neutrino signal outputs are obtained from
collapses with $M_5^{\rm HC}\sim {\cal O}(0.1)$ and
with rotation/magnetic fields.  For these conditions
the average neutrino energy ($\propto (M_5^{\rm HC})^{-0.5}$)
is high, as is the fraction of gravitational binding energy
carried away by neutrinos.
Rotation and/or magnetic fields are always likely to be present
during the collapses.  It is also possible that the lower mass
supermassive progenitors that yield $M_5^{\rm HC}={\cal O}(0.1)$
dominate the population of supermassive objects. The optimal
conditions of neutrino detection, therefore, may in fact
represent most of the supermassive collapses.

Competing with the neutrino signal is the fluctuation of
the background level of ICECUBE, which at the current AMANDA
level should have a standard deviation of $20N_M^{0.5}$
counts in 1 second.  To obtain a signal-to-noise ratio (S/N)
of unity for a collapse at $z=0.15$ with $M_5^{\rm HC}=0.3$
would then require $2.8\times 10^6$ optical modules. If we
take advantage of the time template of the neutrino signal,
we may not require a signal to noise as high as S/N$=1$.  A
S/N of 0.1, for example, would only require $2.8\times 10^4$ 
optical modules for a detection. The rate of such detections
could be of the order of 0.1 to 1 per year, if most of the
supermassive collapses at $z\la 0.2$ have rotating/magnetized
homologous cores with relatively lower masses in the 
supermassive mass spectrum, $M_5^{\rm HC}={\cal O}(0.1)$.

Since to maintain the same S/N, the required $N_M$ is
proportional to $d^{4}$, a chance
detection of a supermassive collapse at $z\la 0.1$ (and hence
$d\la 500$ Mpc) would make the detection scheme even more attractive.
Note also that the number of modules ($N_M$) required scales 
linearly with the duration of the neutrino burst, which we have
taken to be $t_{\rm dyn}$ in the above estimates.
If this duration is only a fraction of $t_{\rm dyn}$
as shown in the example in figure 1, the number of optical
modules can be further reduced.  In addition, if the energy
of neutrino emission during the collapses can exceed $30\%$
of the total gravitational binding energy, $N_M$ will
scale down as the inverse square of this fraction. Therefore, 
the detection potential actually starts at the
$N_M\sim 5\times 10^3$ level, which is currently
envisaged for the expansion of the operating AMANDA
neutrino telescope to the ICECUBE configuration.
An ICECUBE detector with the order of $10^5$
optical modules has already been contemplated
for other science missions, e.g., for searching
for nucleon decay with a sensitivity that cannot
be matched by conventional techniques.

The above calculations have been based on the
detection technology available to AMANDA, without
any change in design of the planned ICECUBE.  This
leaves room for future improvement.
Moreover, if the formation of supermassive black holes
via supermassive object collapse also gives
rise to $\gamma$-ray bursts\cite{fuller98},
or gravitational radiation detectable
in proposed low frequency gravitational wave
detectors such as LISA, we could enhance our
ability to detect the associated neutrino bursts
by knowing when to look. In addition, the combined
neutrino/$\gamma$-ray burst/gravitational wave 
signal would offer a golden chance to explore many
questions in particle physics, astrophysics and 
gravitational physics.

We thank K. Abazajian, G. Burbidge, E. M. Burbidge, K. Griest,
J.~X. Prochaska, and P. Shaver for discussions. X.~S. and
G.~M.~F. are supported by NASA grant NAG5-3062 and NSF
grant PHY98-00980 at UCSD.
F.~H. is supported by DOE grant DE-FG02-95ER40896
and by the Wisconsin Alumni Research Foundation.

\newpage
\begin{table}
\caption{Expected Neutrino Events per Optical Module in
ICECUBE from Collapses at $z=0.15$.\label{table1}}
\begin{tabular}{lcccc}
$M^{\rm HC}\,(10^5M_\odot)$&
Rotation/Magnetic Field&
$\alpha$&
$E_{\nu\bar\nu}/0.5M^{\rm HC}c^2$&
$N_{\rm event}/N_M$\\\tableline
0.3 & No & 0.8 & 7\% & $8\times 10^{-4}$ \\
0.3 & Yes & 9 & 30\%\tablenote{Saturating our imposed limit.}
    & $1.2\times 10^{-2}$ \\
1 & No & 0.01 & 4\% & $1\times 10^{-5}$ \\
1 & Yes & 1.2 & 30\%$^a$ & $5\times 10^{-3}$ \\
3 & Yes & 0.017 & 22\% & $1.7\times 10^{-4}$ \\
\end{tabular}
\end{table}
\newpage
\noindent{\bf Figure Captions:}

\noindent
Figure 1. Solid line: the time profile of the (un-normalized)
neutrino luminosity from the collapse of a non-rotating
non-magnetized spherical supermassive object\cite{shi98}.
Dashed line: the time profile of the neutrino signal in
the ICECUBE detector.  The time axis is in units of
$t_{\rm dyn}$, the dynamic collapse timescale.
\bigskip

\noindent
Figure 2. The relative proper number density of quasars as a
function of redshift, normalized to peak at 1.  Data points
are the measurements of Shaver {\sl et al.}\cite{shaver98}.
The thin solid line is the fit of Shaver {\sl et al.}
\cite{shaver98} to the data.  The dashed 
line is inferred from the quasar catalogue of V\'eron-Cetty and
V\'eron\cite{veron}, without any correction for bias.
The three thick lines represent our adopted values
for the relative quasar number density in the redshift range
$0.1\le z\le 0.2$.

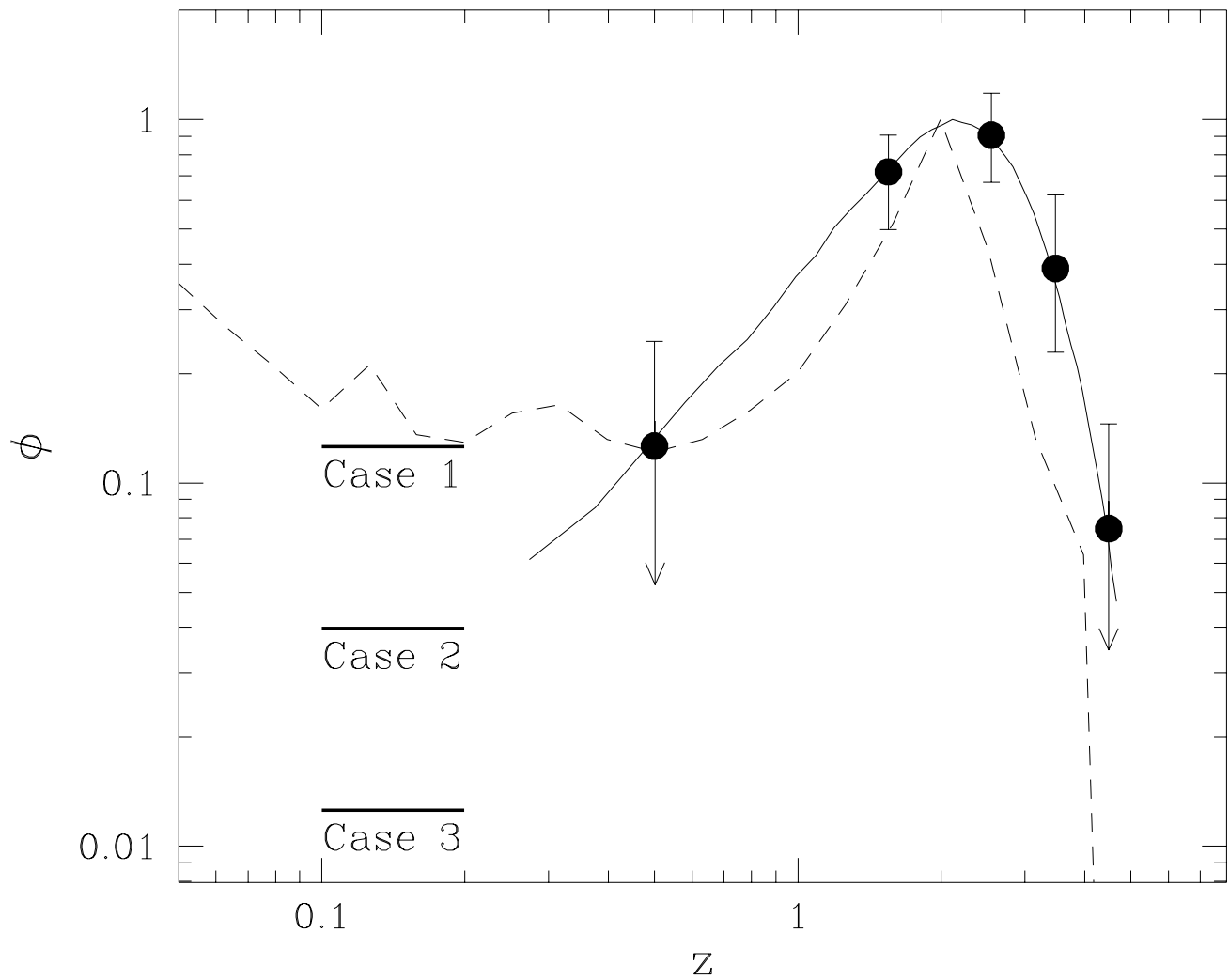
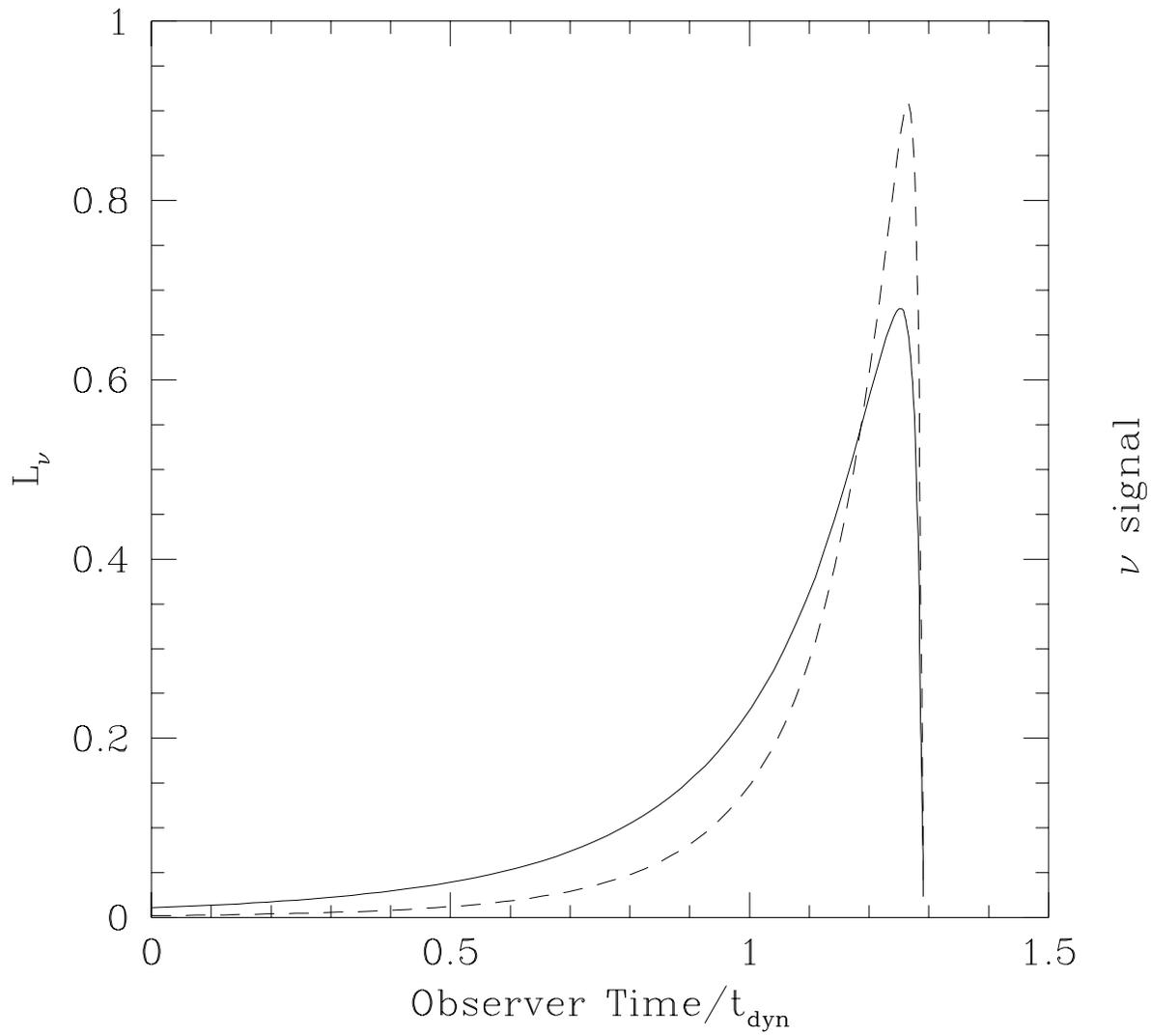
\end{document}